\documentclass[]{aastex}
\usepackage{emulateapj5}
\usepackage{onecolfloat}
\usepackage{graphicx} 
\usepackage{fancyheadings} 
\usepackage{ulem}
\usepackage{rotating}
\usepackage{lscape}

\newcommand{\tskip}{\tablevspace{1pt}}

\newcommand{\ndla}{13}
\newcommand{\snpix}{15}

\newcommand{\delv}{\Delta v}

\newcommand{\lya}{Ly$\alpha$ }

\newcommand{\cm}[1]{\, {\rm cm^{#1}}}
\newcommand{\N}[1]{{N({\rm #1})}}

\newcommand{\sci}[1]{{\rm \; \times \; 10^{#1}}}
\newcommand{\ltk}{\left [ \,}

\newcommand{\rtk}{\, \right  ] }

\newcommand{\smm}{\sum\limits}

\newcommand{\cmma}{\;\;\; ,}

\newcommand{\mkms}{{\rm \; km\;s^{-1}}}

\begin{document}

\twocolumn[%
\submitted{Accepted to the Astrophysical Journal: September 3, 2002}

\title{THE CHEMICAL UNIFORMITY OF HIGH $z$ DLA PROTOGALAXIES}

\author{ JASON X. PROCHASKA\altaffilmark{1}}
\affil{UCO/Lick Observatory}
\affil{University of California, Santa Cruz;
Santa Cruz, CA 95064}
\email{xavier@ucolick.org}

\begin{abstract} 

We investigate chemical abundance variations along the sightlines
through 13 damped \lya systems (DLA).  
We introduce a technique designed to identify abundance variations in 
multiple velocity bins along the sightlines and perform a series of
Monte-Carlo simulations to derive quantitative limits and values. 
The majority of these DLA have very uniform
relative abundances: 95$\%$ c.l.\ upper limits of 0.2~dex dispersion 
along the sightline and best fit values typically $\lesssim 0.1$~dex.  
This surprising result indicates the gas comprising
individual DLA has similar enrichment history, nearly identical
differential depletion, and small abundance variations relating to ionization
corrections.  This uniformity contrasts with stellar
abundance variations observed within the Galaxy and with 
the differences in abundances observed between galaxies of the Local group.  
It also contrasts with variations in 
differential depletion along sightlines through the SMC, LMC, and Milky Way.
The results constrain the process of metal production and dust formation
in the early universe and reflect on processes of
galaxy formation.  
In terms of depletion the observations indicate a very low porosity of 
significantly depleted gas, substantially lower than the filling factor
observed in present-day metal-poor galaxies.
Finally, the observed chemical uniformity
may present a difficult challenge to scenarios which assume individual DLA are
comprised of multiple protogalactic clumps.

\end{abstract}

]

\pagestyle{fancyplain}
\lhead[\fancyplain{}{\thepage}]{\fancyplain{}{PROCHASKA}}
\rhead[\fancyplain{}{THE CHEMICAL UNIFORMITY OF HIGH $z$ DLA PROTOGALAXIES}
]{\fancyplain{}{\thepage}}
\setlength{\headrulewidth=0pt}
\cfoot{}

\altaffiltext{1}{Visiting Astronomer, W.M. Keck Telescope.
The Keck Observatory is a joint facility of the University
of California and the California Institute of Technology.}

\section{INTRODUCTION}
\label{sec-intro}

Studies of chemical abundance ratios in Galactic metal-poor stars
address processes of nucleosynthesis, star formation, and 
ultimately galaxy formation in the early universe.
Of particular importance is revealing variations or trends in the
relative abundances which identify distinct stellar populations,
unique nucleosynthetic sites, and/or provide insight into the
time-scales and channels of star formation.
In very low metallicity ([Fe/H]~$< -3$) Galactic stars, 
for example, observers have identified
important trends in the Fe-peak elements and large variations
in several abundance ratios (e.g.\ Co/Fe, Sr/Fe) which suggest unique
nucleosynthetic processes and inhomogeneous chemical enrichment
\citep[e.g.][]{mcw95,johnson02}.
In contrast, the abundance patterns of the majority of 
Galactic stars with $-2 < $~[Fe/H]~$< -1$ 
are remarkably similar \citep{gratton91,fulbright00}, suggesting
a rapid enrichment phase for the Galactic halo over this metallicity range.
At higher metallicity,  there are important trends in the relative abundances,
in particular the steady decline of the $\alpha$-elements (O,S,Si)
to the Fe-peak, which mark the onset of Type~Ia SN enrichment \citep{tinsley79}.
Comparisons of these relative abundances with extragalactic stellar 
populations help map paths of galaxy formation
\citep[e.g.][]{matteucci01}.
In the few galaxies where detailed studies can be performed
\citep[the SMC and dwarf spheroidals;][]{venn98,shetrone01,bonifacio00}, one
often observes significant differences with stars at the same
metallicity in the Milky Way.  Future studies of entire stellar populations
within these galaxies will help refine their star formation histories.

An examination of relative abundances of gas in local galaxies 
complements these stellar abundance studies.
Sightlines through the Galactic ISM probe gas with a range of physical
conditions (e.g.\  volume density, dust-to-gas ratio, ionization state).
Therefore, the observed {\it gas-phase} abundance ratios along a sightline
may vary by an order of magnitude
as the sightline penetrates clouds arising in various phases of the ISM
\citep[e.g.][]{sav96}.
Differences in nucleosynthetic enrichment or photoionization
are likely to be small such that the variations 
primarily reflect differences in depletion level or dust composition.
The LMC and SMC also exhibit variations in these gas-phase
abundance ratios \citep{welty99,welty01} and together 
the observations impact models for the 
formation and distribution of dust as well as 
the physical state of galactic interstellar media.

Observations of the damped \lya systems (DLA) -- quasar absorption line systems 
with $\N{HI} \geq 2 \sci{20} \cm{-2}$ -- probe the ISM of 
high $z$ galaxies.  Surveys of the DLA reveal chemical
evolution in the early universe \citep{ptt94,ptt97,pw00}, and 
examine the dust depletion,
nucleosynthetic enrichment, and ionization of these protogalaxies
\citep{lu96,vladilo98,pw02,pro02}. 
\cite{pw02} recently reviewed the
chemical abundances of the DLA and emphasized the
uniformity in the relative abundance patterns (e.g.\ Si/Fe, Ni/Fe, Zn/Si)
of their $\approx 30$ systems.  
Even though damped \lya metallicities span over an order of magnitude, 
the systems show very similar abundance ratios
suggesting protogalaxies have common enrichment histories.
The authors do identify a mild trend of increasing Si/Fe and Zn/Fe
with increasing metallicity which is well explained by the effects
of dust depletion, yet even this trend spans only a factor of $\approx 3$
in the ratios.
In addition to the uniformity in abundance patterns from galaxy 
to galaxy, \cite{pw02} 
asserted that this uniformity holds along the sightlines penetrating each
protogalaxy.  That is, the individual 'clouds' comprising metal-line profiles
show the same relative abundances {\it within} a given damped \lya system.
If verified, this would have
important implications for the ISM of high $z$ galaxies and the enrichment
of gas in the early universe.

\cite{pro96} first quantitatively 
investigated variations in the chemical abundances
of a single damped \lya system.
They compared
Zn, Si, Fe, Ni, and Cr ionic column densities along the observed
velocity profiles
using the apparent optical depth method \citep{sav91}.
Their analysis showed that
the chemical abundances were uniform to within 
statistical uncertainties. 
They then argued that the absence of significant variations demonstrates
a low depletion level within this protogalaxy.
More recently, \cite{lopez02} performed the first detailed line-profile analysis 
of the relative abundances for
a different $z>2$ damped system and also found nearly constant
column density ratios among the components comprising the line-profile
solution (see also Petitjean, Srianand, \& Ledoux 2002).  
Together, these studies argue the gas within high $z$
protogalaxies has a similar enrichment history and uniform
differential depletion.

In this paper, we search for variations in the relative abundances 
along the sightlines of \ndla\
damped \lya systems.  
These systems were selected to have high signal-to-noise velocity
profiles with velocity widths exceeding 40~km\,s$^{-1}$.  The latter criterion
ensures a large enough baseline to examine variations within the protogalaxy.
We introduce a quantitative method which can be 
applied to our observations as well as mock spectra derived from
numerical simulations and semi-analytic models
of galaxy formation.  The technique assesses changes 
in the physical conditions and enrichment
histories of the gas within each damped system. In turn, 
our results place valuable constraints on the various morphological models
of the damped \lya systems and, ultimately, scenarios of galaxy formation.

\begin{table*} \footnotesize
\begin{center}
\caption{{\sc OBSERVATIONAL SAMPLE \label{tab:summ}}}
\begin{tabular}{lcccccccccccc}
\tableline
\tableline
QSO &$z_{abs}$ & [M/H]\tablenotemark{a} & $\delv$ &
Transition & [X/Y]$_T$ & $\delta v_{bin}$ &
$N_{b}$ & $\sigma_{bin}$\tablenotemark{b} &
$\chi^2_\nu$ & $\Delta_{sngl}$ & $\Delta_{all}$ & $\Delta_{best}$ \\
& & & (km\,s$^-1$) & Pairs & (dex) & (km\,s$^{-1}$)  & & (dex) & & (dex) & (dex) 
& (dex)\\
\tableline
\tskip
PH957     & 2.309 & --1.46 & 54 & Zn\,II 2026 & 0.22 & 10 & 3 & 0.18 & 1.65 & 0.24 & 0.47 
 & 0.03 \\
          &       &        &    & Ni\,II 1741 \\
Q0201+36  & 2.463 & --0.41 & 202& Si\,II 1808 & 0.45 & 20 & 12& 0.07 & 3.21 & 0.38 & 0.15 
 & 0.09 \\
          &       &        &    & Fe\,II 1608 \\
Q0347--38 & 3.025 & --1.17 & 94 & Si\,II 1808 & 0.50 & 10 & 5 & 0.47 & 11.0 & 0.74 & 0.90 
 & 0.39 \\
          &       &        &    & Fe\,II 1608 \\
Q0458--02 & 2.040 & --1.19 & 86 & Zn\,II 2026 & 0.39 & 20 & 4 & 0.48 & 4.81 & 0.35 & 0.45 
 & 0.15 \\
          &       &        &    & Cr\,II 2056 \\
HS0741+47 & 3.017 & --1.69 & 40 &  S\,II 1259 & 0.24 & 10 & 5 & 0.08 & 2.23 & 0.23 & 0.22 
 & 0.07 \\
          &       &        &    & Fe\,II 1608 \\
PSS0957+33& 4.180 & --1.50 & 346& Si\,II 1304 & 0.37 & 20 & 6 & 0.26 & 5.34 & 0.70 & 0.68 
 & 0.29 \\
          &       &        &    & Fe\,II 1608 \\
Q1223+17  & 2.466 & --1.59 & 94 & Si\,II 1808 & 0.13 & 10 & 10& 0.11 & 1.82 & 0.40 & 0.19 
 & 0.09 \\
          &       &        &    & Ni\,II 1751 \\
Q1331--17 & 1.776 & --1.30 & 70 & Zn\,II 2026 & 0.82 & 10 & 9 & 0.23 & 22.8 & 0.66 & 0.41 
 & 0.24 \\
          &       &        &    & Fe\,II 2374 \\
GB1759+75 & 2.625 & --0.82 & 78 & Si\,II 1808 & 0.35 & 10 & 8 & 0.18 & 2.50 & 0.44 & 0.27 
 & 0.11 \\
          &       &        &    & Ni\,II 1751 \\
Q2206--19 & 1.920 & --0.42 &130 & Si\,II 1808 & 0.40 & 20 & 7 & 0.06 & 1.17 & 0.24 & 0.15 
 & 0.04 \\
          &       &        &    & Fe\,II 1611 \\
Q2230+02  & 1.864 & --0.74 &177 & Si\,II 1808 & 0.23 & 20 & 7 & 0.08 & 1.30 & 0.31 & 0.18  
 & 0.03 \\
          &       &        &    & Ni\,II 1741 \\
Q2231--00 & 2.066 & --0.86 &127 & Si\,II 1808 & 0.28 & 20 & 4 & 0.08 & 0.93 & 0.26 & 0.33 
 & 0.03 \\
          &       &        &    & Ni\,II 1741 \\
Q2359--02 & 2.154 & --1.58 &127 & Si\,II 1526 & 0.32 & 20 & 4 & 0.08 & 0.97 & 0.24 & 0.32  
 & 0.04 \\
          &       &        &    & Fe\,II 1608 \\
\tableline
\end{tabular}
\tablenotetext{a}{Metallicity derived from Zn, Si, or S}
\tablenotetext{b}{Logarithmic RMS dispersion in X/Y}
\tablecomments{$\Delta_{sngl}, \Delta_{all}, \Delta_{best}$ are measures of the
intrinsic abundance variation along each sightline.  See $\S$~3}
\end{center}
\end{table*}

\section{ANALYSIS OF THE OBSERVATIONS}
\label{sec-anly}

To investigate variations in the relative abundances
of the damped systems, we focus on pairs of low-ion
transitions for \ndla\ damped systems from our damped \lya 
database\footnote{http://kinpgin.ucsd.edu/$\sim$hiresdla/} \citep{pro01}.
We selected damped \lya systems with transition pairs that satisfied
the following criteria:
(1) signal-to-noise (S/N) $>$~\snpix\ per pixel;
(2) a velocity width $\delv \geq 40$~km\,s$^{-1}$ 
to provide a base-line for examining abundance variations;
(3) transitions which are not saturated but have peak optical 
depth with $>5\sigma$ significance;
(4) a transition pair with different refractory characteristics,
nucleosynthetic origin, and/or photoionization dependencies.  With respect
to  the
last criterion, the $\alpha$-ions (e.g.\ Si$^+$, S$^+$) are ideal complements
to the Fe-peak ions (e.g.\ Ni$^+$, Fe$^+$).  The former are 
non-refractory or mildly refractory, 
are primarily produced in Type\,II SN and are less ionized
in H\,II regions, while the latter are refractory, primarily produced in 
Type\,Ia SN, and in the case of Fe$^+$ are more highly ionized.
Table~\ref{tab:summ} summarizes the damped \lya systems examined in this
paper and lists the transitions comprising our analysis.
In nearly every case, we observe multiple transitions for 
a given damped \lya system and are confident that the transitions listed
in Table~\ref{tab:summ} are free of line-blending.

We considered two techniques for probing variations in the relative
abundances of a pair of metal-line profiles:
(1) a comparison of the ionic column densities derived 
from the apparent optical depth \citep{sav91}
in a series of velocity bins; and 
(2) a comparison of the column densities derived from Voigt profile
fitting of
individual velocity components. 
The latter approach has the advantage that one presumably isolates
abundances in specific physical components (e.g.\ 'clouds' or 'clumps'),
whereas the velocity bins may contain one or more such components.
There are several weaknesses, however, to the Voigt profile
analysis.  First, identifying the component structure of a metal-line profile
is both time consuming and somewhat arbitrary.  One typically 
introduces enough 'clouds' to produce a line-profile solution that
yields a reduced $\chi_\nu^2 \approx 1$.  Therefore, high S/N profiles invariably
exhibit more components than lower quality data.  Second, 
blended components generally exhibit degeneracies between their column densities.
In order 
to accurately compare ionic column densities one must bin
these blended components and carefully consider their correlated errors.
Third and most important, in cases where a pair of transitions have very
large abundance differences it may be impossible to derive a self-consistent
profile solution.  For example, one ion may show several components which
are not observed in the other ion.  In these cases, one cannot compare
the abundances component by component.  The risk is that this
would bias the results against systems with large abundance variations.

\begin{figure*}[ht]
\begin{center}
\includegraphics[height=5.5in, width=4.5in]{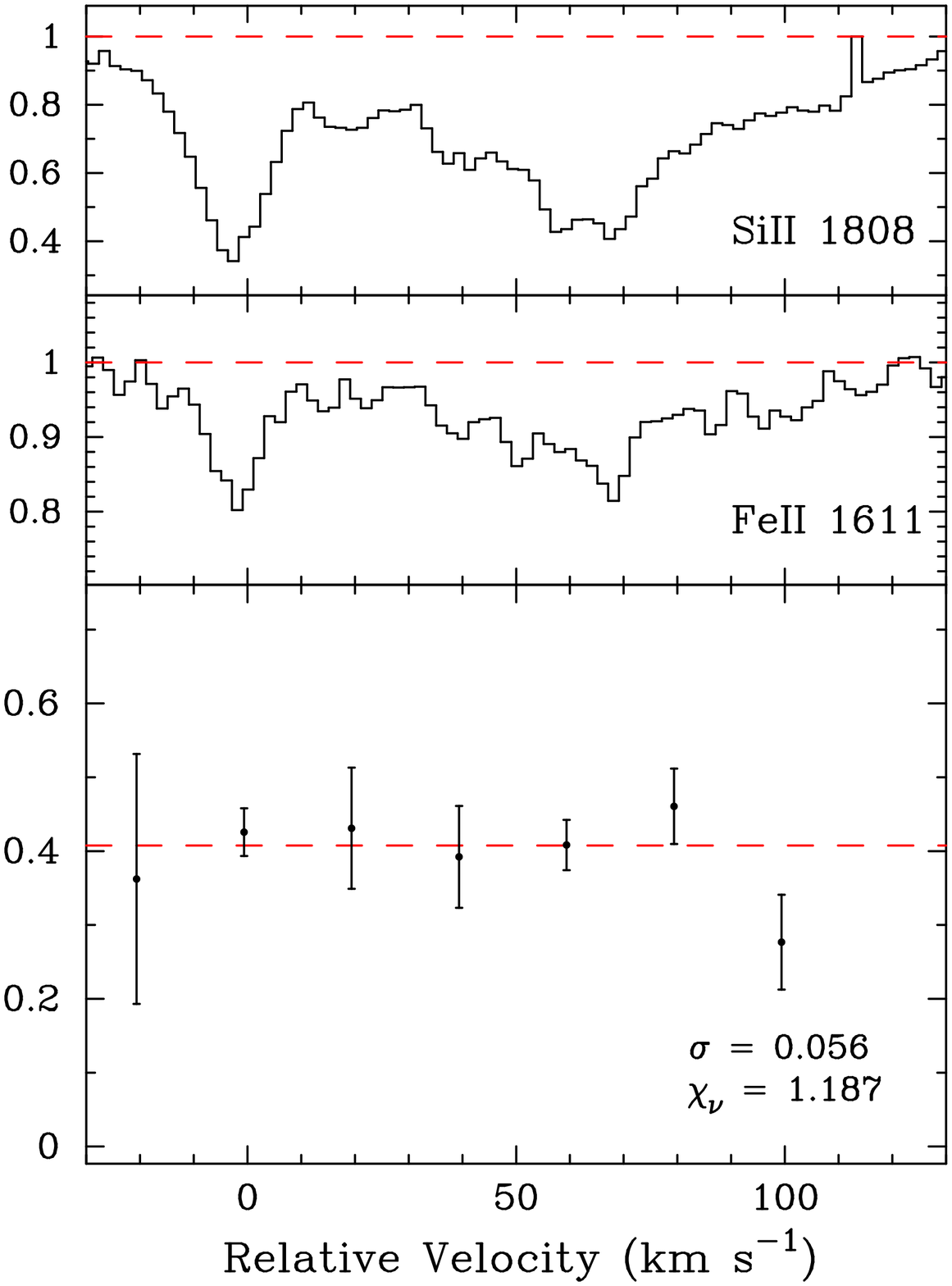}
\caption{The top panels show the velocity profiles for the Si\,II 1808
and Fe\,II 1601 transitions for the damped \lya system at $z=1.92$ toward
Q2206--19.  The lower panel plots the logarithmic ratio of the column 
densities of the two ions measured using the apparent optical depth method
in 20~km\,s$^{-1}$ bins where the dashed line indicates the total
logarithmic ratio.  Deviations from this total value are small for
the 7 bins ($\sigma < 15\%$) and they are consistent with expectation from
statistical error alone.
}
\label{fig:ex}
\end{center}
\end{figure*}

In this paper, we concentrate on the apparent optical depth method.
For many of the sightlines presented here, we have also examined the results
from a Voigt profile analysis and find similar results
\citep[e.g.\ the systems included in an analysis
of the fine structure constant;][]{murphy01}.
After experimenting with the size of velocity bins,
we found that binning the apparent optical depth
over 10~pixels (20~km\,s$^{-1}$) offered the best comprise between
minimizing pixel-to-pixel noise and maximizing velocity resolution.  
For the very highest S/N data, we reduced the 10~pixel bin to 5~pixels.
In all cases, we clipped velocity bins with statistical significance 
less than $2.5 \sigma$, i.e., in regions where there is very little absorption.
At 20~km\,s$^{-1}$ or even 10~km\,s$^{-1}$, it is possible that the velocity bins
include two or more 'clouds' with differing abundances which
are averaged to a single value;  we do not resolve abundance
variations on this small scale.  On the other hand, the typical
separation between the main components comprising the metal-line
profiles is $\gtrsim 20 \mkms$.  Our analysis focuses on
these main components.

\begin{figure*}
\begin{center}
\includegraphics[height=8.5in, width=6.0in]{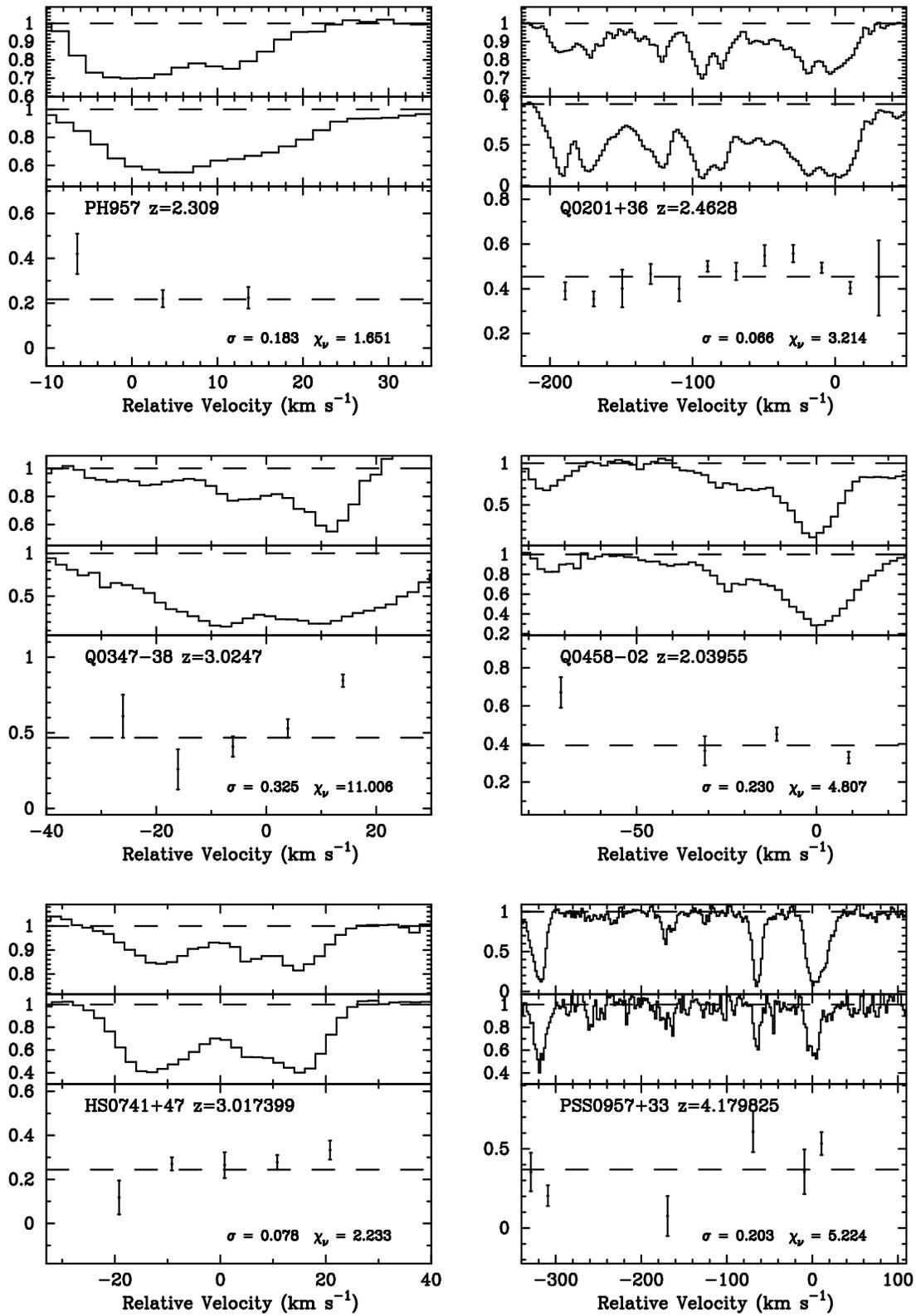}
\caption{
Same as Figure~1 but for the remaining damped \lya systems comprising the
sample.
}
\label{fig:prof}
\end{center}
\end{figure*}

Figure~\ref{fig:ex} presents the results for the damped \lya system
at $z=1.92$ toward Q2206--19 which is roughly representative of the 
full sample (Figure~\ref{fig:prof}).  The upper panels show the 
Si\,II 1808 and
Fe\,II 1611 velocity profiles while the lower panel plots the ratio
$\log_{10}$~(Si$^+$/Fe$^+$) normalized to solar relative abundances 
for the statistically significant velocity bins.
The dashed-line in the lower panel indicates
the ratio of the total column densities derived from  these two profiles.  
We compare the individual bins against this total ratio to examine
departures from uniformity.
The error bars only reflect statistical error;
we do not account for systematic errors like continuum placement.
In the systems analysed in this paper, which have relatively high quality
data, we expect such systematic errors are small.
The damped \lya system presented in Figure~\ref{fig:ex} shows very
small variations over the entire profile.  
The scatter in the relative abundances is $<15\%$ ($\approx 0.05$~dex)
of the total ratio, consistent with the expected statistical error, i.e.,
the reduced $\chi^2_\nu \approx 1$.

\begin{figure*}[ht]
\begin{center}
\includegraphics[height=5.5in, width=4.5in]{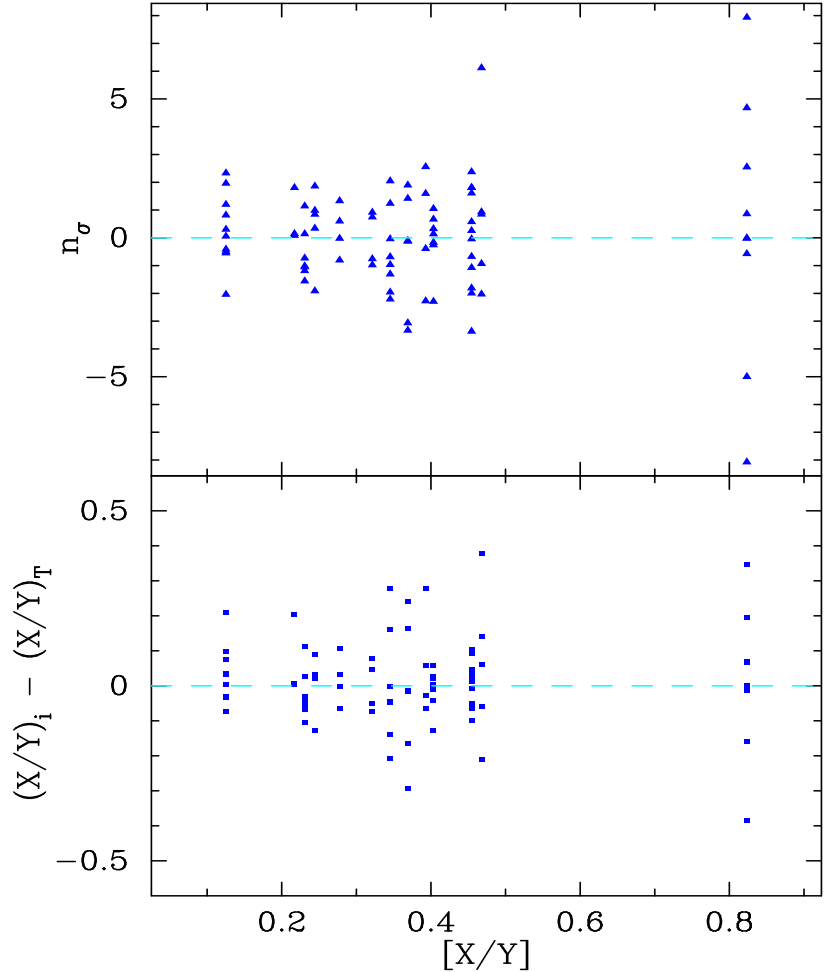}
\caption{
The two panels summarize the (a) statistical significance and
(b) logarithmic magnitude of the deviations from uniformity for all of
the velocity bins of our DLA sample as a function of the integrated
X/Y ratio, [X/Y]$_T$.  Only two systems exhibit variations at greater than 0.3~dex
and the majority of bins have values less than 0.1~dex.
}
\label{fig:all}
\end{center}
\end{figure*}

We have repeated this analysis for the remaining damped systems 
and summarize the results for all of the velocity bins for the \ndla\ DLA
in Figure~\ref{fig:all}.
The top panel shows the
departure from uniformity in numbers of standard deviation $n_\sigma$ 
where $\sigma$ is derived with standard propagation of statistical error. 
The lower panel presents the departures on a logarithmic scale relative
to the total X/Y ratio.
Together the two panels describe the magnitude and
significance of the variations in our sample.
For both panels, the x-axis indicates the total enhancement of X/Y 
relative to solar
on a logarithmic scale [X/Y]$_T$.  Higher [X/Y]$_T$ values indicate systems with larger
nucleosynthetic enhancements, dust depletion, and/or photoionization levels.

Consistent with a qualitative inspection of the damped \lya profiles,
there are only two systems with deviations $>0.3$~dex and only a few others
have statistically significant deviations at greater than $0.2$~dex.
Of the two systems with the largest departures from uniformity, one
is the damped system at $z=1.77$ toward Q1331+17 which shows the
largest [X/Y] value and might have been most expected to show significant
deviations.  
The other system, at $z=3.025$ toward Q0347--38, was 
previously known to have abundance differences among its metal-line
profiles and H$_2$ absorption lines \citep{levshakov02}.
Inspecting Figure~\ref{fig:all}, one infers that the abundance variations
within most damped systems are at levels less than 0.2~dex.  In the following
section, we quantitatively address the maximum variations allowed by the
entire sample and the maximum deviations allowed by individual damped systems.

\begin{figure*}
\begin{center}
\includegraphics[height=8.5in, width=6.0in]{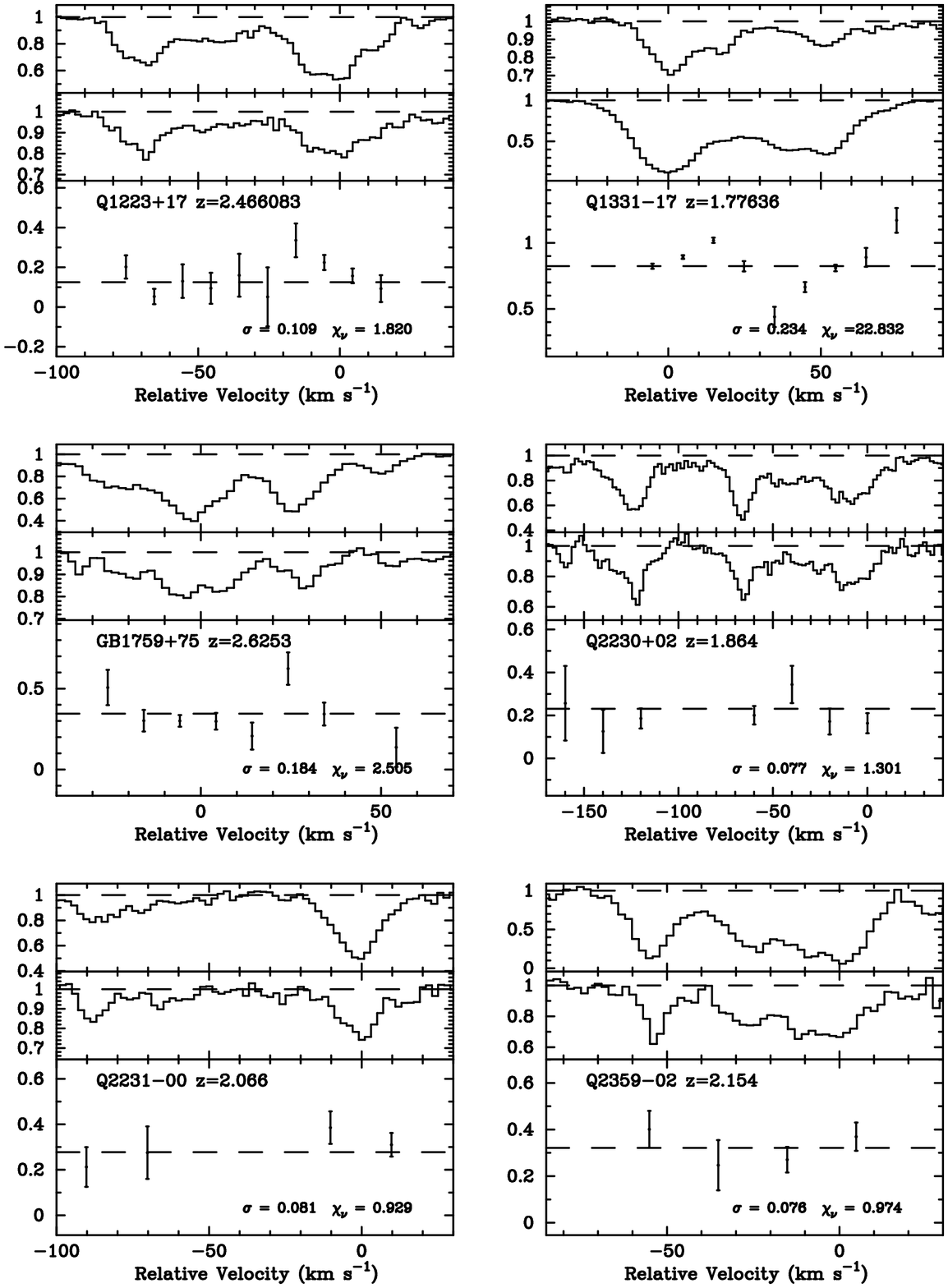}
\end{center}
\end{figure*}

\section{MEASURES OF INTRINSIC VARIATION}
\label{sec-discuss}

To place quantitative constraints on the nucleosynthetic and
differential depletion variations in the DLA, we will now determine the 
maximum abundance variations allowed by our observations.
For a crude measure of the typcial abundance variation, 
we evaluate the maximum component-to-component
variation allowed when considering the {\it entire sample} of 
damped \lya systems together, 
i.e., the maximum intrinsic abundance variation which is
consistent with the complete distribution of observed
deviations (Figure~\ref{fig:all}).
To this end, we place limits on the intrinsic abundance dispersion $\delta$
which is the combined variation due to nucleosynthetic enrichment,
differential depletion, and photoionization effects.
Specifically,
$\delta$ defines a uniform distribution of intrinsic abundance variations 
for the $N_{b}$ velocity bins comprising a single DLA.

\begin{figure*}[ht]
\begin{center}
\includegraphics[height=5.5in, width=4.0in, angle=90]{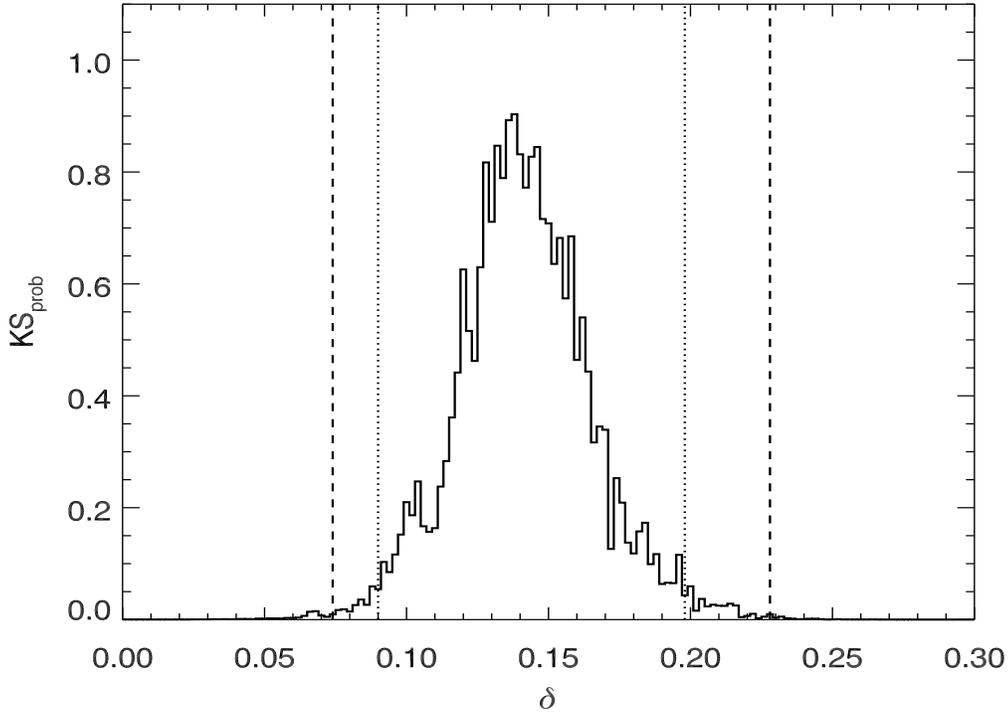}
\caption{
Confidence level plot for a series of Monte-Carlo simulations assuming
a range of intrinsic abundance variation $\delta$.  
The dotted lines indicate the approximate 95$\%$ c.l. values and the
dashed lines refer to $99 \%$ c.l.
These simulations indicate that the 
damped \lya systems do not allow intrinsic variations greater than 0.25~dex
at the 99$\%$ c.l.  At the same time, the observations do require an
intrinsic variation at $>0.05$~dex.
}
\label{fig:contour}
\end{center}
\end{figure*}

For $\delta$ values ranging from 0 to 0.3~dex, we perform a series
of Monte-Carlo analyses.  In each Monte-Carlo run, we 
simulate 1000 damped systems each with
$N_{b}$ velocity bins.  For every velocity bin, we randomly 
draw four quantities: 
(1) the statistical uncertainty $\sigma$ in the abundance ratio, drawn from the 
observed distribution of uncertainties for all bins of all of the
damped systems in our sample;
(2) the contribution to the observed ratio from this statistical
uncertainty assuming a Gaussian distribution $\Delta_{error}$ with
mean of zero and variance $\sigma^2$;
(3) a intrinsic abundance variation $\Delta_{abnd}$ drawn 
uniformly\footnote{Note that we
assume uniform distributions for the abundance deviations throughout 
this paper.
If we adopted a normal distribution, we would place even
tighter constraints on the abundance variations.}
from the interval $\Delta_{abnd} \, \epsilon \, [-\delta,\delta]$;
and
(4) a weighting factor $w_i$ for each bin drawn uniformly from 1 to 10 in
order to simulate the observed variability in component strengths.
We then calculate the total ratio: 

\begin{equation}
T = \log_{10} \ltk \frac{\smm_{i=1}^{N_{b}} 
w_i \cdot 10^{\Delta_i}}{\smm w_i}  \cmma
\rtk
\end{equation}
where $\Delta_i = \Delta_{abnd} + \Delta_{error}$ which corresponds
to the [X/Y]$-$[X/Y]$_T$ values displayed in Figure~\ref{fig:all}.
For each velocity bin in every damped system, we determine the
statistical significance 
\begin{equation}
n_\sigma \equiv (T-\Delta_i)/\Delta_{error}
\end{equation}
of the variation and for the 1000 trials
determine a distribution of $n_\sigma$ values.  This Monte-Carlo distribution
of $n_\sigma$ values is then compared against the observed distribution
(top panel of Figure~\ref{fig:all}) with a two-sided Kolmogorov-Smirnov (K-S)
test.  The K-S statistical test focuses on the median of
two distributions and the K-S probability gives the likelihood that
two distributions could have been drawn from the same parent population\
\citep{press92}.

In Figure~\ref{fig:contour} we present the results of our Monte-Carlo
simulations as a confidence-level plot for a range of $\delta$ values.
The 99$\%$ c.l. limits for $\delta$ are 0.07 and 0.23~dex.
If a single abundance process dominates the abundance variations, these
are the limits on that process.  If multiple processes contribute
(e.g.\ differential depletion and nucleosynthetic enrichment), then
the limits on each process are even tighter.
Note, these conclusions are insensitive 
to the value of $N_{b}$ for $N_{b} > 2$.
The results in Figure~\ref{fig:contour} quantitatively support
the characteristic of uniformity readily observed in the metal-line profiles.
If depletion and nucleosynthetic variations are present in the
damped systems, they are present at very small levels in the majority
of protogalaxies.  While the simulations imply strict upper limits to
the magnitude of abundance variations, Figure~\ref{fig:contour}
also reveals that the observations are inconsistent with zero abundance
variations.  At the 95$\%$~c.l., we find that $\delta$
must exceed $\approx$0.05~dex to match the observations\footnote{
It is possible that we are slightly overestimating this lower limit
(and the upper limits) because we have not included systematic error
in our analysis, but we expect this would contribute at $< 0.05$~dex.}.
Therefore, the data do require some intrinsic abundance variations.

Impressed by the absence of abundance variations in
the detailed profiles as well as the integrated [X/Y]$_T$ values,
we posed the following hypothesis:  
Are the abundance ratios within individual damped systems consistent 
with being drawn from
a single population described by the distribution of [X/Y]$_T$ values?  
To address this hypothesis, 
we performed a similar Monte-Carlo simulation where
the [X/Y] values for each velocity bin
are drawn from the observed [X/Y]$_T$ for all of 
the damped \lya systems in our database \citep[over 30 DLA;][]{pro01}, i.e., 
the variations $\Delta_i$ are due to differences in [X/Y]$_T$
among the observed damped \lya systems.  We simulated 1000 damped systems
with $N_{b}=4$ bins each and found
a KS probability $P_{KS} = 49\%$ suggesting the dispersion in the
component-by-component abundances of the DLA is consistent with the 
dispersion in the integrated [X/Y]$_T$ values:  
{\it The damped systems exhibit a 'scale-free dispersion' in their
relative abundances.}
This trait did not have to hold for these protogalaxies.
One could surmise a scenario where
the gas 'clouds' comprising a single damped system would have the same
physical properties and abundances ratios but their [X/Y]$_T$ values
would differ by greater than 1~dex from system to system.  
Conversely, the gas comprising a 
single damped system could exhibit a range of depletion levels and
nucleosynthetic enrichment which average out to a very similar [X/Y]$_T$ ratio
in all damped systems.  In our current sample, however, we find that both
the gas within a single damped system and the systems together have
similar dispersion in the relative abundances.

The results presented in Figure~\ref{fig:contour} provide an estimate
of the overall variations in the damped \lya systems. 
One can perform a similar set of Monte-Carlo simulations to 
investigate the deviations allowed within each damped system and
then examine trends with other
physical properties of the damped systems 
(e.g.\ [X/Y]$_T$, as in Figure~\ref{fig:all}).
We will consider three measures of the uniformity within individual
damped systems.  With one measure, we study an extreme scenario
where all of the variation arises from a single velocity bin.
We calculate, $\Delta_{sngl}$, the minimum [X/Y] variation in 
a single velocity bin which
gives $\chi^2_{M-C} > \chi^2_{Obs}$ in over $95\%$ of the trials in 
the Monte-Carlo analysis.
The two other measures assume deviations in all of the velocity bins
with values drawn from a uniform distribution.  We calculate
(i) $\Delta_{all}$, the minimum variation 
which when applied to every velocity bin gives
$\chi^2_{M-C} > \chi^2_{Obs}$ in over $95\%$ of the trials; and
(ii) $\Delta_{best}$, the variation which applied to each component
has the highest probability of yielding $\chi^2 = \chi^2_{Obs} \pm 10\%$.
We believe the $\Delta_{all}$ values are the most realistic upper limits
to abundance variations and that $\Delta_{best}$ 
reflects the most likely value.

For each damped system, we take the observed number of velocity bins and
adopt the measured statistical errors.  To simplify the analysis, we
parameterize all of the sources of intrinsic abundance variations into a single
term as above.  Columns 11-13 of Table~\ref{tab:summ} present the results
from a series of Monte-Carlo simulations comprised of 1000~trials for
each damped system. 
It is important to stress that $\Delta_{sngl}$ and $\Delta_{all}$ each
has a systematic behavior related to the number of bins observed.
Generally, systems with lower $N_{bin}$ will have statistically lower
$\Delta_{sngl}$ values and higher $\Delta_{all}$ values, while the
opposite holds for systems with $N_{bin} > 5$.  This behavior is simply
related to Poissonian statistics.
In general, the derived $\Delta_{sngl}$ and $\Delta_{all}$ values corroborate
the picture revealed by Figure~\ref{fig:contour}, i.e.,  the upper limits
on the observed variations in individual DLA are between 0.2 and 0.4~dex.
In several systems -- especially those with higher S/N --
the upper limits are below 0.2~dex at the 95$\%$~c.l. 
and $\Delta_{best} < 0.1$~dex.   Within these systems, it is difficult
to accommodate any significant variation in the intrinsic abundances.

\section{IMPLICATIONS}

\subsection{Nucleosynthetic Enrichment and Dust Depletion}

The results presented in the previous section demonstrate that the
majority of damped \lya systems have relative ionic column abundances which
are uniform to better than 0.2~dex.  In fact, the majority of components
show departures $<0.1$~dex.
This analysis reveals an important characteristic of high $z$ galaxies: 
the gas 'clouds' which comprise these protogalaxies apparently have
very similar physical properties and nucleosynthetic enrichment histories.  

\begin{table}[hb]\footnotesize
\begin{center}
\caption{{\sc DUST SUMMARY\label{tab:dust}}}
\begin{tabular}{lccccccccccc}
\tableline
\tableline
Gal & Phase\tablenotemark{a} &
D(Zn) & D(Si)\tablenotemark{b} & D(Fe) &
[Zn/Fe] & [Si/Fe] & Ref  \\
\tableline
MW & warm halo & 0.0   & --0.3 & --0.6 & +0.6 & +0.3 & 1 \\
MW & warm disk & 0.0   & --0.4 & --1.4 & +1.4 & +1.0 & 1\\
MW & cold disk & --0.1 & --1.3 & --2.2 & +2.1 & +0.9 & 1\\
LMC& warm halo & 0.0   &--0.2  & --0.4 & +0.4 & +0.2 & 3\\
LMC& warm disk & 0.0   &--0.1  & --0.9 & +0.9 & +0.8 & 3\\
LMC& cold disk & 0.0   &--0.4  & --1.2 & +1.2 & +0.8 & 3\\
SMC& warm halo & 0.0   &  0.0  & --0.6 & +0.6 & +0.6 & 2\\
SMC& warm disk & 0.0   &  0.0  & --1.4 & +1.4 & +1.4 & 2\\
SMC& cold disk & 0.0   &  0.0  & --2.2 & +2.2 & +2.2 & 2\\
\tableline
\end{tabular}
\tablenotetext{a}{Crude phase description, largely based on depletion level}
\tablenotetext{b}{Depletion factor (e.g.\ [X/Zn])}
\tablerefs{Key to References -- 1: \cite{sav96}; \\
2: \cite{welty99}; 3: \cite{welty01}}
\end{center}
\end{table}

In several respects this uniformity is unexpected.  
Consider first the variations one might predict from nucleosynthetic enrichment.
We stress that the metallicities of the individual gas 'clouds' comprising
a damped \lya system are unknown because the H\,I line-profiles are
unresolved.  Variations in metallicity 
greater than a factor of 10 are possible {\it if not probable}.
The components may have very different enrichment levels 
and could be expected to show different nucleosynthetic patterns.
The magnitude of variations in the nucleosynthetic patterns, however, would
depend on the actual range of metallicities and the detailed star formation
histories.  
While extremely metal-poor Galactic halo stars exhibit significant
variations in their relative abundances 
\citep[e.g.\ Co/Ni, Zn/Fe:][]{mcw95,johnson02}, the
majority of Galactic stars with [Fe/H]~$=-2$ to $-1$~dex
have nearly uniform relative abundances \citep{gratton91,fulbright00}.
There are, however, a number of stars and globular clusters in
this metallicity range which show significantly lower $\alpha$/Fe ratios
\citep[e.g.\ Ru 106, Pal~12, 
$\omega$~Cen, etc.;][]{brown97,nissen97,smith00,stephens02}.
Although these stars may have a distinct origin from the bulk of the Galactic
halo stars (e.g.\ accreted from other galaxies), 
they reveal nucleosynthetic variations at all metallicity.

\begin{figure*}[ht]
\begin{center}
\includegraphics[height=5.5in, width=4.5in]{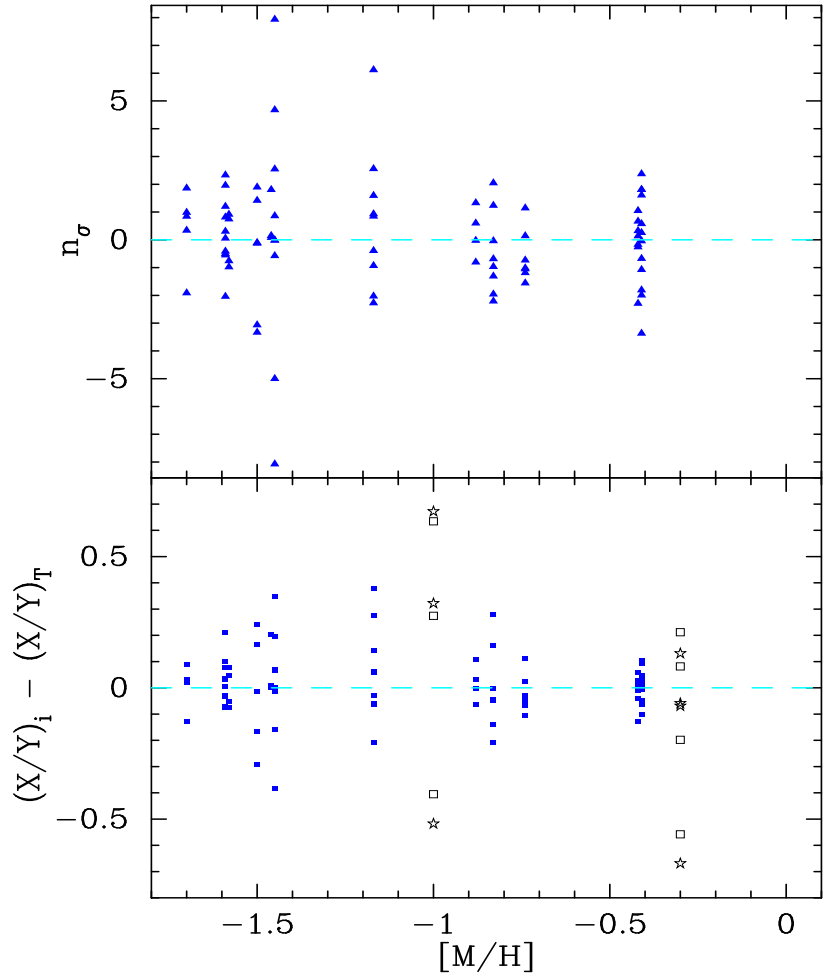}
\caption{
Same as Figure~\ref{fig:all} except the deviations are plotted as a function
of DLA metallicity.  For comparison in the lower panel, 
we show the abundance variations 
measured in the LMC (at [M/H]~$= -0.3$) and SMC (at [M/H]~$= -1$) for
Si/Fe (squares) and Zn/Fe (stars).
}
\label{fig:mtl}
\end{center}
\end{figure*}

This point is further emphasized by the differences 
between the Milky Way abundances and other local galaxies 
\citep[e.g.\ LMC, SMC, dSph galaxies:][]{venn98,shetrone01}.
In the SMC,  stars with metallicity [Fe/H]~$= -0.7$
have solar Si/Fe ratios \citep{venn98} in stark contrast with 
values of [Si/Fe]~$>+0.2$ observed at the same metallicity in the Milky Way.  
Similarly, the few measurements of dwarf spheroidal galaxies
\citep{shetrone01,bonifacio00,shetrone02} show a dispersion in Si/Fe of 0.25~dex
at metallicity between 1/10 and 1/100 solar.
Altogether, one observes a range in $\alpha$/Fe of at least 0.3~dex
both within a given galaxy and from galaxy to galaxy.
One might have predicted a similar dispersion in the majority of high
$z$ protogalaxies.  On the other hand, it is important to note that
\cite{kobulnicky97} found only small deviations in the H\,II regions 
throughout NGC~1569.  They concluded that the by-products of star formation
within this galaxy reside in a hot 10$^6$~K phase which is unmixed with
the general ISM.  If this process is important at high $z$, it would
impact the abundance variations within DLA.

In addition to the implications for nucleosynthetic enrichment in the damped
systems, our observations impact the nature
and prevalence of dust in high $z$ protogalaxies.
Table~\ref{tab:dust} summarizes the typical depletion levels and 
gas-phase abundance ratios for various interstellar phases of the
Milky Way, LMC, and SMC \citep{sav96,welty97,welty99,welty01}.  
Because the [Si/Fe] and
[Zn/Fe] ratios vary significantly from phase to phase, most sightlines
penetrating these galaxies show variations in excess of 0.5~dex.
For Si/Fe, the situation is complicated because Si is refractory, e.g., 
one observes roughly the same Si/Fe ratio in the Milky Way warm and
cold disk gas even though the depletion levels differ by $> 0.5$~dex 
(Table~\ref{tab:dust}).
Nevertheless, most sightlines through the Galaxy and Magellanic Clouds
show large variations in Si/Fe.
This is in stark contrast with our observations of the damped \lya systems.

When comparing the damped systems against these galaxies, it
is important to keep in mind that 
even the SMC is more metal-rich than most damped systems.  
If dust formation is very sensitive to gas metallicity, 
one might expect variations in
differential depletion for only the most metal-rich damped systems.
Interestingly,
we observe no correlation between the magnitude of abundance variations and
metallicity.  Figure~\ref{fig:mtl} plots the deviations from uniformity
and their significance as a function of the gas metallicity (either Si/H,
Zn/H or S/H) including a set of values for the LMC and SMC 
\citep{welty99,welty01}. 
Note that the systems with the highest metallicity 
have among the most uniform abundances in our sample. 
It appears that the ISM of the majority of DLA -- 
{\it independent of metallicity} -- is significantly different from modern
galaxies.  This begs an obvious but puzzling question:
How is a galaxy like the SMC different from the majority of DLA
such that it regularly shows highly depleted gas?
We also emphasize that the low values of $\Delta_{sngl}$ 
indicate it is rare in the damped systems to have even
a single, highly depleted 'cloud'.  
If highly depleted gas exists at high redshift, its filling factor
is very small in most protogalaxies.  

Finally, independent of expectations from abundance variations 
observed in local galaxies,
there is a long standing debate on whether
the damped \lya abundances reveal an underlying Type~Ia or
Type~II SN enrichment pattern \citep[e.g.][]{lu96,vladilo98,pw02,vladilo02}.
These two patterns have $\alpha$/Fe ratios differing 
by $\approx 0.3$~dex and imply very different enrichment histories
\citep[e.g.][]{lindner99}.
The source of this debate is the difficulty in disentangling the nucleosynthetic
pattern from the depletion patterns of the gas-phase abundances.
Although this matter remains unsettled, it is possible -- if not likely -- 
that some damped systems are dominated by Type~Ia SN enrichment with
solar $\alpha$/Fe ratios while others exhibit the $\alpha$-enriched
patterns associated with Type~II SN.  Most relevant to the current
discussion, however, is the prospect that  
both processes contribute to the abundances of individual protogalaxies.
The observed chemical uniformity of the damped \lya systems
places tight constraints on possible nucleosynthetic variations. 
For example, scenarios where the majority of gas has nearly solar
Si/Fe or Zn/Fe ratios with a single component enhanced by 0.3~dex are ruled
out in over half of the DLA.

\subsection{The Physical Nature of the DLA}

To fully appreciate the implications of the DLA abundance uniformity,
one would like to interpret the results in terms of the physical nature of the
damped systems.
Unfortunately, the morphology of the damped \lya systems remains an 
open and difficult question.  
While the low-ion kinematics are remarkably well described by
a thick, rotating disk \citep{pw97,pw98}, the observations also
permit scenarios within the CDM
hierarchical cosmology that describe DLA as multiple merging 'clumps' bound to
individual dark matter halos \citep{hae98,mcd99,maller01}.
Furthermore,
several authors have also claimed that the damped \lya systems might be
explained by outflows from SN winds \citep{nulsen98,schaye01}. 
The difficulty in distinguishing between these scenarios stems from the
limited information provided by the 'pinhole' nature of quasar absorption
line studies. 
Although a statistical sample of systems can confidently rule out many
proposed scenarios \citep{pw97}, 
a given sightline cannot uniquely express the galaxy's morphology
and even a sample of sightlines is limited.
In the following, we reflect on several of the
leading scenarios but avoid outflow scenarios because these models
have only been crudely developed and it remains unclear whether
they can successfully account for the velocity fields
of most DLA.  Nevertheless, many of the conclusions for
the other models apply to the wind scenarios.

Assume, first, that the DLA primarily arise from 
gas 'clouds' within a single, gravitationally bound system.
In terms of nucleosynthetic variations, 
the observed uniformity might be easily accommodated within this scenario.  
In particular, one can assume that the majority of gas and stars
comprising the protogalaxy have had a similar chemical enrichment history.
There might be regions of local star formation which would exhibit
a dispersion in nucleosynthetic enrichment, but these regions would
have small filling factor and the by-products may not be directly
dispersed into the general ISM \citep[e.g.][]{kobulnicky97}.
For larger, more massive systems one might require that the
system is well mixed, i.e., the dynamical time is short compared to
the age of the system. 
At $z \sim 2$, the galactic dynamical time is $t_{dyn} < 50$~Myr
assuming $t_{dyn} \sim 2R_d/v_c$ and $R_d = \lambda r_{200}/\sqrt{2}$
where $\lambda$ is the spin parameter and $r_{200}$ is the virial radius
\citep{mmw98}.  Therefore, it is reasonable to expect that the nucleosynthetic
enrichment within a single protogalaxy would not lead to variations 
greater than 0.1~dex and possibly much less.

In contrast with nucleosynthetic variations, the absence of differential
depletion variations is striking.
The low $\Delta_{sngl}$ values for most damped systems
indicate it is rare for sightlines to penetrate even a single, highly
depleted gas 'cloud'.  
This is in stark contrast to the LMC, SMC, and Milky Way where one routinely
intersects clouds with a broad range of depletion levels.
It appears that the damped systems are comprised almost exclusively
of either undepleted gas or gas with warm halo depletion patterns.
If gas with high levels of depletion exist
at high $z$, its porosity is very small, at least within the 
surface density contour defining a damped \lya system.
Perhaps highly depleted gas exists only in the inner, densest regions of
a protogalaxy which has very small cross-section relative to the outer
layers of gas.  
This observational constraint may have important implications for
the detailed nature of star formation in young galaxies.

Within the CDM cosmogony, the scenarios which best reproduce the 
damped \lya kinematics of larger $\delv$ systems are those 
where quasar sightlines
penetrate multiple 'clumps' or satellites within a given dark
matter halo.  
These satellites
represent the remnants of mergers between the present dark matter halo and a
previous (now consumed) halo.  
In this scenario, the uniformity of abundance ratios within the damped \lya 
abundances constrains the relative abundances of all of the satellites
comprising a damped system.  
Presumably, these satellites have unique enrichment histories and
unique physical characteristics, e.g., gas fraction, molecular
gas content, density, velocity fields.
In terms of nucleosynthesis, this implies the satellites share a similar
enrichment history.

On theoretical grounds, the observed $\alpha$/Fe ratio corresponds to the
ratio of Type~II to Type~Ia SN and, therefore, crudely assesses the
rate of star formation \citep[e.g.][]{tinsley79}.  As described above,
galaxies in the local universe exhibit a range of $\alpha$/Fe ratios at the range
of metallicities observed for the damped \lya systems.
Altogether, one observes a variation in $\alpha$/Fe of $\approx$0.3~dex
both within a given galaxy and from galaxy to galaxy. 
The accreted satellites which comprise a damped system in these
CDM scenarios might be expected to have a similar range of metallicity, unique
star formation histories, and therefore exhibit a range in $\alpha$/Fe.
The constraints our observations place on these nucleosynthetic histories
depend on the distribution of $\alpha$/Fe values.  
The observations are inconsistent at the 95$\%$ c.l., for example, 
with a bimodal distribution
of $\alpha$/Fe ratios with values of 0 and +0.3~dex.
On the other hand, 
only a few profiles are inconsistent with a scenario where all of 
satellites have $\alpha$/Fe values ranging uniformly from 0 to +0.3~dex
(i.e.\ $\Delta_{all} < 0.15$).  
We stress, however, that the majority of $\Delta_{best}$ values are $<0.1$~dex
indicating that even this distribution is not favored..
Higher S/N data might lower the 95$\%$ c.l. limits to 0.1~dex 
and pose a true challenge to multiple clump scenarios, especially as 
chemical evolution models achieve greater sophistication.

Similar to the 'single system' models, the constraints on dust depletion
variations are strong.
With the exception of a few damped systems (notably Q1331+17, Q0347--38),
the observations require that the 'clumps' have very similar differential
depletion.  In other words, 
most of the gas comprising the individual satellites in these CDM models
must have a very low porosity of gas with highly depleted levels.
From the results in previous papers \citep[e.g.][]{ptt97,pw02}, 
we knew the integrated depletion levels in the damped systems is both 
small and uniform.   We now appreciate that the components comprising
individual damped systems also are largely undepleted and with very
similar differential depletion.  

\section{SUMMARY AND CONCLUDING REMARKS}

We have performed a quantitative analysis of chemical abundance variations
along the sightlines through 13 damped \lya systems.  These systems each
have velocity width $\delv > 40 \mkms$, i.e., large enough base-line
to examine chemical uniformity.
We performed several sets of Monte-Carlo simulations to place upper limits
on relative abundance variations. To our surprise,
the majority of DLA exhibit a high degree of chemical uniformity. In most
cases, the dispersion in intrinsic abundances -- the combined effects of
all abundance variations -- is less than 0.2~dex and
the best values are generally $<0.1$~dex.

For the ratios examined in this paper, 
there are three potential sources of intrinsic abundance
variation: (1) nucleosynthetic enrichment; (2) differential depletion;
and (3) photoionization corrections.
In terms of depletion,
the observed DLA uniformity contrasts with variations in differential depletion 
observed along sightlines through the
SMC, LMC, and Milky Way. 
Although the latter galaxies have higher metallicity than most DLA, 
the difference in depletion variations 
is not explained solely by differences in metallicity; there is  
no correlation between abundance variations and metallicity in the DLA
(Figure~\ref{fig:mtl}).  
For an unknown reason or reasons (perhaps related to dust destruction mechanisms),
the nature of dust at high $z$ is qualitatively different from the local universe.  
Considering nucleosynthetic processes, the
DLA abundances are more uniform than the dispersion in nucleosynthetic
enrichment of the Milky Way as traced by stellar abundances.
Furthermore, one observes a greater dispersion between stars in various galaxies
within the local group than that observed in the gas of the DLA.
Finally, photoionization calculations suggest
that Si$^+$/Fe$^+$ can vary by over 0.2~dex between a neutral
gas and a highly ionized gas \citep{howk99,vladilo01,pro02}.  Our observations
suggest that the cross-section of H\,II regions or similarly ionized gas
is small for galaxies with $\N{HI} \geq 2 \sci{20} \cm{-2}$.  


The chemical uniformity of the DLA poses an important constraint on the
nature of high $z$ protogalaxies.  We have argued that 'single-system'
scenarios might reasonably account for
the observed abundance invariance.  Central to this conclusion, however,
is that these protogalaxies have a small filling factor of highly depleted
gas.  If star formation is linked to dust-depleted molecular clouds, 
then our results indicate the gas relevant to star formation encompasses
a very small cross-section. This conclusion is consistent with the low fraction of DLA
showing molecular gas \citep{petit00}.
In contrast with the single-system models, we contend the observed uniformity
presents a unique challenge to the multiple-clump scenarios favored by CDM.  
These protogalactic clumps or 'satellites' do not share a common
gas reservoir and should have unique physical characteristics
(density, metallicity) and enrichment histories.
It remains to be demonstrated whether these clumps might express very
similar differential depletion and nucleosynthetic enrichment patterns.

Before concluding, we wish to comment on several future observational
efforts which could improve upon this paper.
First, we emphasize that the results for many of the DLA in the current
sample are limited by signal-to-noise.  
Follow-up observations of the brighter quasars might
reveal uniformity at the 0.01~dex level and place qualitatively tighter
constraints on processes of metal production and dust formation.  
Similarly, higher resolution data would enable an 
investigation in velocity bins of a few km\,s$^{-1}$.  
Second, we would like to repeat our analysis using Zn and Si separately
to better isolate the effects of depletion and nucleosynthesis.
Variations in the Zn/Si ratio have been observed in at least one low $z$
DLA \citep{pettini00} and we would like to examine similar variations 
at high $z$.
Third, it is also important to examine 
abundance ratios which relate to physical processes separate from the
Type~Ia vs.\ Type~II enrichment and differential depletion emphasized in
this paper.  For example, a study on the variations of the 
N/$\alpha$ ratio within DLA would impact our understanding of star
formation timescales and possibly the universality of the initial mass function 
\citep[e.g.][]{pro02b}.
Similarly, because the Ar$^0$/Si$^+$ or O$^0$/Si$^+$ ratios are particularly 
sensitive to photoionization, an analysis of these ions would help reveal
processes of ionization within the DLA. 
Finally, a comparison of the C\,II$^*$ fine-structure line with Si and Fe
resonance transitions bears on the nature of the protogalactic ISM
and ultimately star formation rates \citep{wolfe02}.
Together these observations would help reveal the detailed physical
properties of the gas comprising DLA protogalaxies.

\acknowledgments

The author wishes to recognize and acknowledge the very significant cultural
role and reverence that the summit of Mauna Kea has always had within the
indigenous Hawaiian community.  We are most fortunate to have the
opportunity to conduct observations from this mountain.
The author acknowledges E. Gawiser and M. Murphy who were 
instrumental in starting this project.  This paper also significantly benefited from
discussions with D. Welty.  Finally, the author thanks A. Wolfe, M. Pettini,
and an anonymous referee for helpful comments.
This work was partially supported by NASA through a Hubble Fellowship
grant HF-01142.01-A awarded by STScI to JXP.

\end{document}